# A tutorial on group effective connectivity analysis, part 1: first level analysis with DCM for fMRI

Peter Zeidman*[1], Amirhossein Jafarian[1], Nadège Corbin[1], Mohamed L. Seghier[2], Adeel Razi[3,1], Cathy J. Price[1], Karl J. Friston[1]

[1] Wellcome Centre for Human Neuroimaging
12 Queen Square
London
WC1N 3AR

[2] Cognitive Neuroimaging Unit,
ECAE,
Abu Dhabi,
UAE

[3] Monash Institute of Cognitive & Clinical Neuroscience
Monash Biomedical Imaging
18 Innovation Walk
Monash University
Clayton, VIC, 3800
Australia

* Corresponding author

**Abstract**

Dynamic Causal Modelling (DCM) is the predominant method for inferring effective connectivity from neuroimaging data. In the 15 years since its introduction, the neural models and statistical routines in DCM have developed in parallel, driven by the needs of researchers in cognitive and clinical neuroscience. In this tutorial, we step through an exemplar fMRI analysis in detail, reviewing the current implementation of DCM and demonstrating recent developments in group-level connectivity analysis. In the first part of the tutorial (current paper), we focus on issues specific to DCM for fMRI, unpacking the relevant theory and highlighting practical considerations. In particular, we clarify the assumptions (i.e., priors) used in DCM for fMRI and how to interpret the model parameters. This tutorial is accompanied by all the necessary data and instructions to reproduce the analyses using the SPM software. In the second part (in a companion paper), we move from subject-level to group-level modelling using the Parametric Empirical Bayes framework, and illustrate how to test for commonalities and differences in effective connectivity across subjects, based on imaging data from any modality.

# Contents

# 1   Introduction

Neural models enable us to make inferences about brain circuitry using downstream measurements such as functional magnetic resonance imaging (fMRI). Just as the behaviour of a gas can be described by kinetics equations, which do not require knowing the position of every particle, so neural models can capture the mean activity of large of numbers of neurons in a patch of brain tissue (Deco et al., 2008). A common application of these models in neuroimaging is to assess *effective connectivity* – the directed causal influences among brain regions – or more simply the *effect* of one region on another. This characterisation can be distinguished from the analysis of *functional connectivity*, which concerns statistical dependencies (e.g., the correlation or transfer entropy) between measurements, and *structural connectivity*, which concerns the physical architecture of the brain in terms of white matter tracts and synaptic connections. Effective connectivity cannot typically be observed directly, so models are used to traverse multiple spatial and temporal scales: the microscopic activity of neural populations, the meso- or macroscopic resolution of measurements (for example, LFP, EEG, MEG, ECoG or functional MRI) and population-level effects that are apt for characterising individual subjects.

Dynamic Causal Modelling (DCM) is a framework for specifying models of effective connectivity among brain regions, estimating their parameters and testing hypotheses. It is primarily used in human neuroimaging, but it has also successfully been applied with a range of species including rodents (Papadopoulou et al., 2017) and zebrafish (Rosch et al., 2017). A DCM *forward (generative) model* can be conceptualized as a procedure that generates neuroimaging timeseries from the underlying causes (e.g., neural fluctuations and connection strengths). The generated timeseries depend on the model's *parameters*, which generally have some useful interpretation; for example, a parameter may represent the strength of a particular neural connection. Having specified a forward model, one can then simulate data under different models (e.g. with different connectivity architectures), and ask which simulation best characterises the observed data. Practically, this is done in two stages: first, model *inversion* (i.e., estimation) is the process of finding the parameters that offer the best trade-off between accuracy (the fit of the predicted timeseries to the data) and the complexity of the model (how far the parameters had to move from their prior values to explain the data). This trade-off between accuracy and complexity is quantified by the *model evidence*. In the second stage, hypotheses or architectures are tested by comparing the evidence for different models (e.g. with different network architectures), either at the single-subject or the group level. These two stages are known as Bayesian model *inversion* and *comparison*, respectively. To evaluate the evidence for a model one needs to average over the unknown parameters, which means model inversion is usually needed prior to model comparison. This averaging or marginalisation explains why *model evidence* is sometimes called the *marginal likelihood* of a model.

A variety of biologically informed forward models have been implemented for DCM. These range from simple mathematical descriptions of the gross causal influences among brain regions (Friston et al., 2003) to detailed models of cortical columns, which require temporally rich data afforded by electromagnetic recordings (Moran et al., 2013). In the context of fMRI, the objective of DCM is to explain the interactions among neural populations that show experimental effects. In other words, having identified *where* in the brain task-related effects are localised – usually using a mass-univariate (SPM) analysis – DCM is used to ask *how* those effects came about, in terms of (changes in) the underlying neural circuitry. Figure 1 illustrates the forward model typically used with task-based fMRI experiments. Experimental stimuli drive a neural network model, which predicts the resulting change in neural activity over time. Neural activity is tuned by a vector of parameters $\boldsymbol{\theta}^{(n)}$, which include the strengths of connections and the extent to which the connections are influenced by experimental conditions. The generated neural activity drives a model of neurovascular coupling and haemodynamics, which predicts the resulting change in blood volume and deoxyhaemoglobin level, tuned by the haemodynamic parameters $\boldsymbol{\theta}^{(h)}$. The final part of the model predicts the fMRI timeseries – including noise (e.g. due to thermal variations in the scanner) – one would expect to measure, given the neural activity and haemodynamics, which is configured by parameters $\boldsymbol{\theta}^{(\epsilon)}$. By specifying this forward model and estimating the parameters $\boldsymbol{\theta} = \left(\boldsymbol{\theta}^{(n)}, \boldsymbol{\theta}^{(h)}, \boldsymbol{\theta}^{(\epsilon)}\right)$, the variance in the observed timeseries is partitioned into neural, haemodynamic and noise contributions.

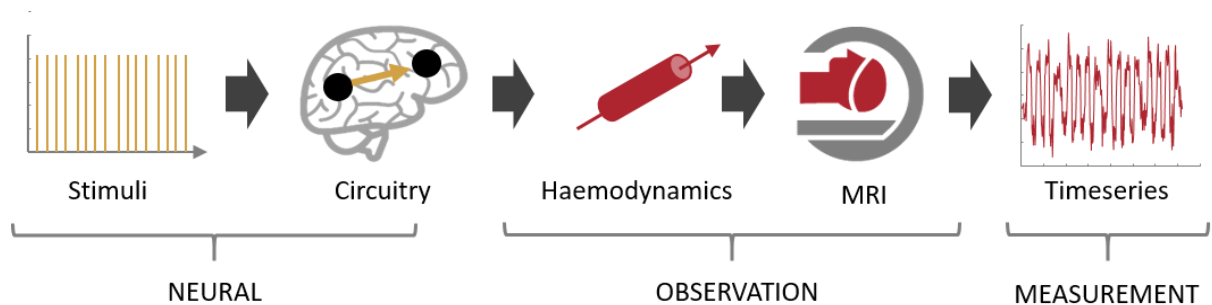

Figure 1. The forward (generative) model in DCM for fMRI. This is split into three parts: neural, observation (subsuming neurovascular, haemodynamic, BOLD signal components) and measurement (the addition of observation noise). The neural model is driven by experimental stimuli, specified as short events (delta functions). The resulting neural activity causes a change in blood flow (haemodynamics), mediated by neurovascular coupling, and consequently the generation of the BOLD signal. The addition of observation noise gives the fMRI timeseries. Image credits: Image credits: "Brain image" by parkjisun and "CT Scan" by Vectors Market from the Noun Project.

To illustrate the methodology – and detail the theory behind it – we analysed data from a previously published fMRI study on the laterality of semantic processing (Seghier et al., 2011). Language is typically thought to be left lateralised; however, the right hemisphere also responds in language tasks. This experiment asked how the left and right frontal lobes interact during semantic (relative to perceptual) processing. We do not attempt to offer any new insights into laterality or semantic processing here; rather we use these data to work through each step of a DCM analysis in detail. In the main text, we survey the current implementation and the specific models used for fMRI. In the appendices, we provide additional technical detail on the models and their implementation in the SPM software package. We hope this tutorial-style overview of the theory will complement and expand on previous reviews and tutorials on DCM (Stephan, 2004; Seghier et al., 2010; Stephan et al., 2010; Kahan and Foltynie, 2013). The example data and a step-by-step guide to running these analyses can be found at https://github.com/pzeidman/dcm-peb-example .

## 2   Notation

Vectors are denoted by lower case letters in bold italics ($\boldsymbol{a}$) and matrices by upper case letters in bold italics ($\boldsymbol{A}$). Other variables and function names are written in plain italics ($f$). The dot symbol ($\cdot$) on its own means multiplication and when positioned above a variable (e.g. $\dot{z}$) denotes the derivative of a variable with respect to time. An element in row $m$ and column $n$ of matrix $\boldsymbol{A}$ is denoted by $A_{mn}$. All variables and their dimensions are listed in Table 1. To help associate methods with their implementation in the SPM software (http://www.fil.ion.ucl.ac.uk/spm/software/), MATLAB function names are provided in bold text, such as (**spm_dcm_fit.m**).

## 3   Experimental design

DCM is a hypothesis-driven approach, the success of which depends on having an efficient experimental design. First, hypotheses need to be clearly articulated, which may relate to effects at the within-subject level, the between-subject level, or both. Here, the within-subject hypothesis was that processing the meaning of familiar words (i.e., their semantic content) would induce greater responses in left frontal cortex than right frontal cortex. The between-subject hypothesis was that this difference in hemispheric responses, quantified by the 'Laterality Index' (LI), would vary across subjects and could be explained by the strength of specific connections.

An efficient experimental design, at the within-subject level, typically involves independently varying at least two experimental factors. Commonly, one factor will be a manipulation of the stimuli that *drive* neural responses, and another factor will be a manipulation of the task demands or context that *modulates* these responses. The distinction between driving and modulatory effects will be made explicit in the DCM analysis that follows. Here, we had two independent factors at the within-subject level: stimulus type (*Words* or *Pictures*) and task (*Semantic* or *Perceptual* reasoning), forming a balanced factorial design with four experimental conditions (words + semantic, words + perceptual, pictures + semantic, pictures + perceptual). An interaction between these two factors was hypothesised; namely, a greater response to words than picture stimuli, specifically in the context of the semantic task. Here, we will test for this interaction using DCM.

# 4 Region selection and fMRI timeseries extraction

DCM is used to model the connectivity between brain regions of interest (ROIs), and the criteria for defining ROIs varies across studies. For resting state experiments, there are no experimental effects, so ROIs are typically selected using an Independent Components Analysis (ICA), or using stereotaxic co-ordinates or masks from meta-analyses or the literature. For task-based experiments, such as that used here, ROIs are usually selected based on an initial mass-univariate SPM analysis, where the objective of DCM is to find the simplest possible functional wiring diagram that accounts for the results of the SPM analysis. Seghier et al. (2011) evaluated an SPM contrast for the main effect of task and identified four ROIs in frontal cortex: 1) left ventral, lvF, 2) left dorsal, ldF, 3) right ventral, rvF, 4) right dorsal, rdF. Relevant timeseries were extracted, pre-processed and summarised within each ROI by their first principal component (see *Appendix 1: Timeseries extraction*). Figure 2 illustrates the experimental timing and timeseries from an example subject.

# 5 Neural model specification

DCM partitions the variability in a subject's timeseries into neural and non-neural (i.e. haemodynamic and noise) sources. This necessitates a two-part model, which can be written generically as follows:

$$\dot{z} = f(z, U, \theta^{(n)})$$
$$y = g(z, \theta^{(h)}) + X_0 \beta_0 + \epsilon$$
(1)

The first line describes the change in neural activity due to experimental manipulations. The level of neural activity within all the modelled brain regions are encoded by a vector $z$. These are the *hidden states*, which cannot be directly observed using fMRI. The function $f$ is the neural model (i.e., a description of neuronal dynamics), which specifies how the change in neural activity over time $\dot{z}$ is caused by experimental stimuli $U$, current state $z$, and connectivity parameters $\theta^{(n)}$. On the second line of Equation 1, the function $g$ is the haemodynamic model, which specifies the biophysical processes that transform neural activity $z$ into the Blood Oxygen-Level Dependent (BOLD) response with parameters $\theta^{(h)}$. The remainder of the second line comprises the measurement or noise part of the model. A General Linear Model (GLM) with design matrix $X_0$ and parameters $\beta_0$ captures known uninteresting effects such as the mean of the signal. Finally, zero-mean I.I.D. observation noise $\epsilon$ is modelled, the variance of which is estimated from the data (see *Appendix 2: Observation noise specification*).

The choice of neural $f$ and observation model $g$ depends on the desired level of biological realism and the type of data available. Here, we used the default neural and haemodynamic models for fMRI data, first introduced in Friston et al. (2003), which captures slow emergent dynamics that arise from coupled neural populations. Models are specified by answering a series of questions (Q1-Q8) in the DCM software, which are described in the following sections.

## 5.1 Input specification

The first question, when specifying a DCM, is which experimental conditions to include, to be specified as columns in $U$ (Figure 2, left). We had three conditions – Task (all semantic decision trials), Pictures (the subset of trials in which subjects made semantic judgements based on picture stimuli) and Words (the subset of trials with written stimuli). Trials of the perceptual control task and incorrect trials were not modelled and so formed the implicit baseline. Having selected these conditions, SPM imports the onset times of the trials automatically from the initial GLM analysis. (Note that when trials have a positive duration – i.e. they are blocks – the corresponding columns of $U$ have value one during stimulus presentation and zero when stimuli are absent. In the special case where all the trials are events with zero duration, $U$ is scaled by the number of time bins per second.)

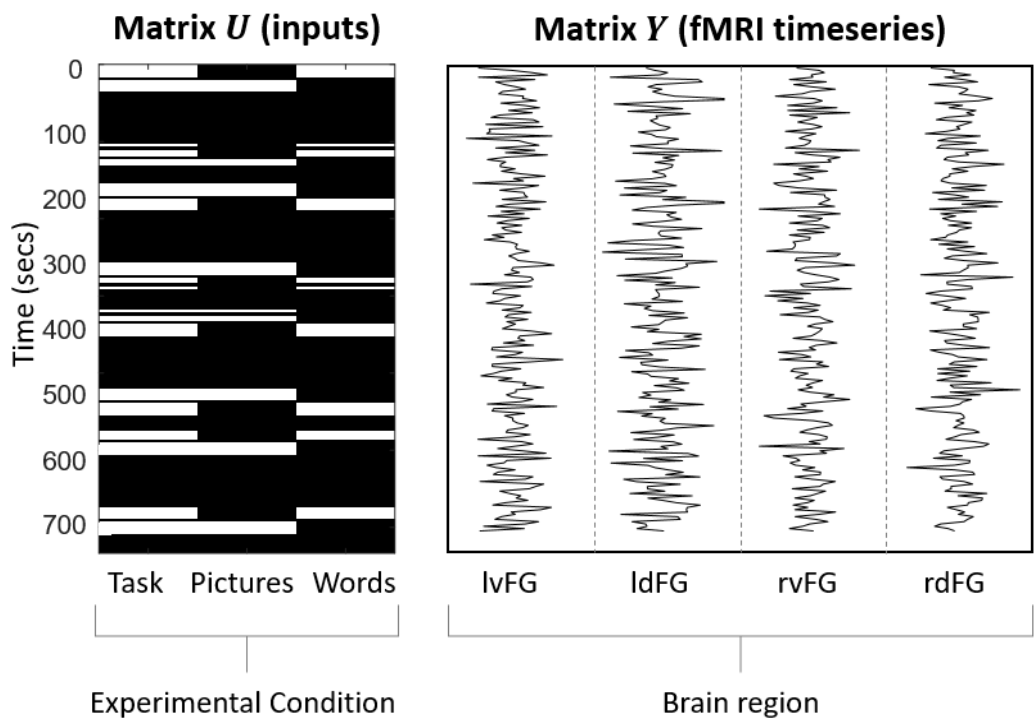

Figure 2 Prerequisites for DCM analysis of task fMRI data: the design ($U$) and data ($Y$). Left: Experimental inputs $U$. White areas indicate times during the experiment when experimental stimuli were shown to the subject. There were three conditions: 'Task' comprised all semantic decision trials, 'Pictures' and 'Words' comprised the subset of trials for each condition. Right: fMRI timeseries $Y$ for each of the four brain regions to be modelled from a typical subject. These are concatenated vertically to give data vector $y$ specified in Equation 1.

## 5.2 Slice timing

Whereas the neural state $z$ is continuous, fMRI data are discrete, with a volume acquired every 3.6 seconds in our data (the repetition time, TR). A strategy is therefore needed to align the acquisition of the fMRI data to the model. Most fMRI data are acquired in sequential slices, meaning that measurements from different brain regions (located in different slices) will be separated in time. DCM has a slice timing model (Kiebel et al., 2007) that enables the acquisition time of each region to be modelled, which may particularly benefit models with widely spaced brain regions. However, this assumes that we know the time at which each slice was acquired, which is generally not the case – because the brain is rotated and deformed during spatial normalisation. Furthermore, MRI sequences that do not acquire slices in sequential order (e.g. interleaved or multi-band sequences) would not be properly represented by the slice timing model. If in doubt, the typical approach is to minimise slice timing effects by using the middle slice of the volume. Here, we set the slice timing model to use the last slice of the volume (3.6s for all regions) to be consistent with the original publication of these data.

## 5.3 Bilinear or nonlinear

The third question when specifying the DCM is which neural model to use; i.e., how to approximate function $f$ in Equation 1. The default 'bilinear' neural model in DCM for fMRI (**spm_fx_fmri.m**) uses a *Taylor approximation* to capture the effective connectivity among brain regions, and the change in effective connectivity due to experimental inputs. This model was later extended (Stephan et al., 2008) to include a nonlinear term, enabling brain regions to modulate the effective connectivity between other brain regions. Here, we did not need to consider nonlinear effects, so we selected the default bilinear model. The bilinear model captures the change in neural activity per unit time $\dot{z}$ according to:

$$\dot{\boldsymbol{z}} = \boldsymbol{J}\boldsymbol{z} + \boldsymbol{C}\boldsymbol{u}(t)$$

$$\boldsymbol{J} = \left(\boldsymbol{A} + \sum_k \boldsymbol{B}^{(k)} \boldsymbol{u}_{\boldsymbol{k}}(t)\right) \quad (2)$$

$$\boldsymbol{u}(t) = \boldsymbol{U}_{t,:}^T$$

$$\boldsymbol{u}_{\boldsymbol{k}}(t) = \boldsymbol{U}_{t,k}$$

The first line says that neural response $\dot{z}$ depends on connectivity matrix $\boldsymbol{J}$. The columns of this matrix are the outgoing connections and the rows are the incoming connections, so element $J_{mn}$ is the strength of the connection from region $n$ to region $m$ (Under a Taylor approximation, this is also the Jacobian – the partial derivative of neural activity in region $m$ with respect to region $n$). Parameter matrix $\boldsymbol{C}$ is the sensitivity of each region to driving inputs, where element $C_{pq}$ is the sensitivity of region $p$ to driving input from experimental condition $q$. This is multiplied by $\boldsymbol{u}(t)$, the row of $\boldsymbol{U}$ corresponding to all the experimental inputs at time $t$.

The second line of Equation 2 specifies the connectivity matrix $\boldsymbol{J}$, which is configured by two sets of parameters: $\boldsymbol{A}$ and $\boldsymbol{B}$. Parameter matrix $\boldsymbol{A}$ specifies the average or baseline effective connectivity (see *Q6: Centre input*) and $\boldsymbol{B}^{(k)}$ specifies the modulation of effective connectivity due to experimental condition $k = 1 \dots K$. Each matrix $\boldsymbol{B}^{(k)}$ is multiplied by experimental inputs $\boldsymbol{u}_{\boldsymbol{k}}(t)$ relating to condition $k$ at time $t$. In this experiment, we had three $B$-matrices corresponding to $K = 3$ experimental conditions or inputs: *Task* (the onsets of all trials), *Pictures* (blocks in which the stimuli were pictures) and *Words* (blocks in which the stimuli were words). For the derivation of this model, see *Appendix 3: Derivation of the fMRI neural model*, and for more detail on the units and interpretation of the parameters, see *Appendix 4: The neural parameters*.

Importantly, each region in this model is equipped with an inhibitory self-connection, specified by the elements on the leading diagonal of the average connectivity matrix ($\boldsymbol{A}$) and modulatory input matrices ($\boldsymbol{B}^{(k)}$). These parameters control the self-inhibition in each region, or equivalently, their gain or sensitivity to inputs. Biologically, they can be interpreted as controlling the region's excitatory-inhibitory balance, mediated by the interaction of pyramidal cells and inhibitory interneurons (cf. Bastos et al., 2012). These parameters are negative and preclude run-away excitation in the network. This is implemented by splitting the average connectivity matrix ($A$) and modulatory input matrices ($B^{(k)}$) into two parts: intrinsic within-region self-inhibition ($A_I, B_I$) and extrinsic between-region connectivity ($A_E, B_E$). These parts are combined as follows:

$$J = \underbrace{-0.5 \cdot \exp(A_I) \cdot \exp\left(\sum_k B_I^{(k)} u_k(t)\right)}_{\text{Intrinsic (self-inhibition)}} + \underbrace{\left(A_E + \sum_k B_E^{(k)} u_k(t)\right)}_{\text{Extrinsic (between-region)}} \quad (3)$$

where $-0.5 Hz$ is the default strength of the self-connections. $A_I$ and $B_I^{(k)}$ are diagonal matrices, i.e.

$$A_I = \begin{bmatrix} A_{I\,1} & 0 & 0 & \cdots \\ 0 & A_{I\,2} & 0 & \cdots \\ 0 & 0 & A_{I\,3} & \cdots \\ \vdots & \vdots & \vdots & \ddots \end{bmatrix}, B_I^{(k)} = \begin{bmatrix} B_{I\,1}^{(k)} & 0 & 0 & \cdots \\ 0 & B_{I\,2}^{(k)} & 0 & \cdots \\ 0 & 0 & B_{I\,3}^{(k)} & \cdots \\ \vdots & \vdots & \vdots & \ddots \end{bmatrix} \quad (4)$$

and $A_E$ and $B_E^{(k)}$ are off-diagonal matrices as follows:

$$A_E = \begin{bmatrix} 0 & A_{E\,1,2} & A_{E\,1,3} & \cdots \\ A_{E\,2,1} & 0 & A_{E\,2,3} & \cdots \\ A_{E\,3,1} & A_{E\,3,2} & 0 & \cdots \\ \vdots & \vdots & \vdots & \ddots \end{bmatrix}, B_E^{(k)} = \begin{bmatrix} 0 & B_{E\,1,2}^{(k)} & B_{E\,1,3}^{(k)} & \cdots \\ B_{E\,2,1}^{(k)} & 0 & B_{E\,2,3}^{(k)} & \cdots \\ B_{E\,3,1}^{(k)} & B_{E\,3,2}^{(k)} & 0 & \cdots \\ \vdots & \vdots & \vdots & \ddots \end{bmatrix} \quad (5)$$

Equations 3 to 5 specify the same model as in Equation 2, except the self-connections are constrained to be negative. The self-connections $A_I$ and $B_I^{(k)}$ are unitless log scaling parameters. This furnishes them with a simple interpretation: the more positive the self-connection parameter, the more inhibited the region, and so the less it will respond to inputs from the network. Matrices $A_E$ and $B_E^{(k)}$ are the extrinsic connectivity among regions, in units of *Hz*, because they are rates of change. For example, $A_{E\,3,1}$ is the strength of the connection from region 1 to region 3, or equivalently the rate of change in region 3 effected by region 1. If it is positive, then the connection is excitatory (region 1 increases activity in region 3) and if it is negative then the connection is inhibitory. Similarly, $B_{E\,3,1}^{(k)}$ is the increase or decrease in connectivity from region 1 to region 3 due to experimental condition $k$.

In summary, the neural model in DCM for fMRI captures directed interactions between brain regions, with connection strengths encoded in matrices of parameters. Matrices $A_I$ and $B_I^{(k)}$ are the self-connections, which are unitless log scaling parameters. Matrices $A_E$ and $B_E^{(k)}$ are the between-region connections, in units of Hz. Care needs to be taken, therefore, to correctly report the different units of each type of parameter. In the software implementation of this model in SPM (**spm_fx_fmri.m**), the diagonal elements of the connectivity matrices are the self-connections and the off-diagonal elements are the between-region connections. Having elected to use this bilinear model, we were next asked how the activity in each brain region should be modelled.

## 5.4 States per region

The 'one-state' bilinear DCM for fMRI model, described above, represents the level of activity of each brain region $i$ at time $t$ as a single number $z_i(t)$. A subsequent development was two-state DCM (Marreiros et al., 2008) that generates richer neural dynamics, by modelling each brain region as a pair of excitatory and inhibitory neural populations. This has been used, for example, for modelling changes to the motor cortico-striato-thalamic pathway in Parkinson's disease (Kahan et al., 2014). The two-state model requires the use of positivity and negativity constraints on all connections, which needs to be taken into account when interpreting the results (for details, see https://en.wikibooks.org/wiki/SPM/Two_State_DCM). Here, for simplicity, we selected the one-state DCM.

## 5.5 Stochastic effects

The model described in equations 2-5 is deterministic, meaning that the experimental stimuli drive all the neural dynamics. Stochastic DCM (Li et al., 2011) estimates time-varying fluctuations on both neural activity (hidden states) and the measurements. This means that stochastic DCM can be used to model resting state fMRI as well as task-based studies where endogenous fluctuations are important. However, stochastic DCM poses a challenging model estimation problem, as both the connectivity parameters and trajectory of the hidden states need to be inferred. For resting state fMRI, a more recent technology (DCM for Cross-Spectral Densities, (Friston et al., 2014)) offers a simpler and more efficient solution, by modelling the data in the frequency domain (see Section 5.7). By modelling the data features in terms of spectral power, the stochastic fluctuations above become spectral components that are much easier to parameterise and estimate. Here, we elected not to include stochastic effects.

## 5.6 Centre input

The next question is whether to mean-centre input matrix $U$. If experimental input is mean-centred, then the parameters in matrix $A$ represent the average effective connectivity across experimental conditions and modulatory parameters $B^{(k)}$ add to or subtract from this average. If $U$ is not mean-centred, then $A$ is the effective connectivity of the unmodelled implicit baseline (akin to the intercept of a linear model), onto which each modulatory input adds or subtracts. Mean-centring can improve the model evidence, by enabling the connectivity parameters to stay closer to their prior expectation (of zero) during model inversion. Furthermore, it ensures that excursions from baseline activity are reduced; thereby eluding nonlinear regimes of the haemodynamic model. Finally, mean centring also affords the matrix $A$ a simpler interpretation. Here, we chose to mean-centre the inputs, giving positive values in $U$ when stimuli were presented and negative values elsewhere.

## 5.7 Timeseries or Cross-Spectral Density (CSD)

DCM for Cross Spectral Densities (CSD) is used for modelling fMRI data in the frequency domain, rather than the time domain, by fitting second order statistics like the cross-spectral density of the time series. This provides an efficient method for analysing resting state data (Friston et al., 2014). It uses the same neural model as described above, but without modulatory inputs, as it is assumed that the connection strengths remain the same throughout the acquisition. Unlike stochastic DCM, this method does not try to model the neural state fluctuations in the time domain. By fitting data features in the frequency domain, estimation is significantly quicker, more efficient, and more sensitive to group differences (Razi et al., 2015). Here, we chose to fit timeseries rather CSD, because we were interested in condition specific, time-varying connectivity due to the task.

## 5.8 Connections

Having selected the form of the model, the next step is to configure it by specifying which parameters should be switched on (i.e., informed by the data) and which should be switched off (fixed at their prior expectation of zero). It is this sparsity structure that defines the architecture or model in question. Figure 3 illustrates the network architecture we specified for each subject's DCM (**spm_dcm_specify.m**). We will refer to this as the 'full model', because all parameters of interest were switched on. Extrinsic or between-region connectivity parameters (matrix $A_E$) were enabled between dorsal and ventral frontal regions in each hemisphere, and between homologous regions across hemispheres. Heterotopic connections were switched off, in line with previous findings: see the discussion in Seghier et al. (2011).

DCM distinguishes two types of experimental input: *driving* and *modulatory*. Driving inputs are usually brief events that 'ping' specific regions in the neural network at the onset of each stimulus. The resulting change in neural activity reverberates around the network. Modulatory inputs up- or down-regulate specific connections and represent the context in which the stimuli were presented. They are typically modelled as blocks (box-car functions). This stage of the model specification asks which experimental conditions should be driving inputs and which should be modulatory inputs. We set Task (the onset of all *Semantic* trials) as the driving input to all regions (matrix $C$) and we set the context of being in *Pictures* blocks or *Words* blocks as modulatory inputs on the self-connection of each region (the diagonal elements of matrices $B_I^{(2)}$ and $B_I^{(3)}$ respectively). Limiting modulatory effects to the self-connections, rather than including the between-region connections, added biological interpretability (as changes in the excitatory-inhibitory balance of each region) and generally improves parameter identifiability.

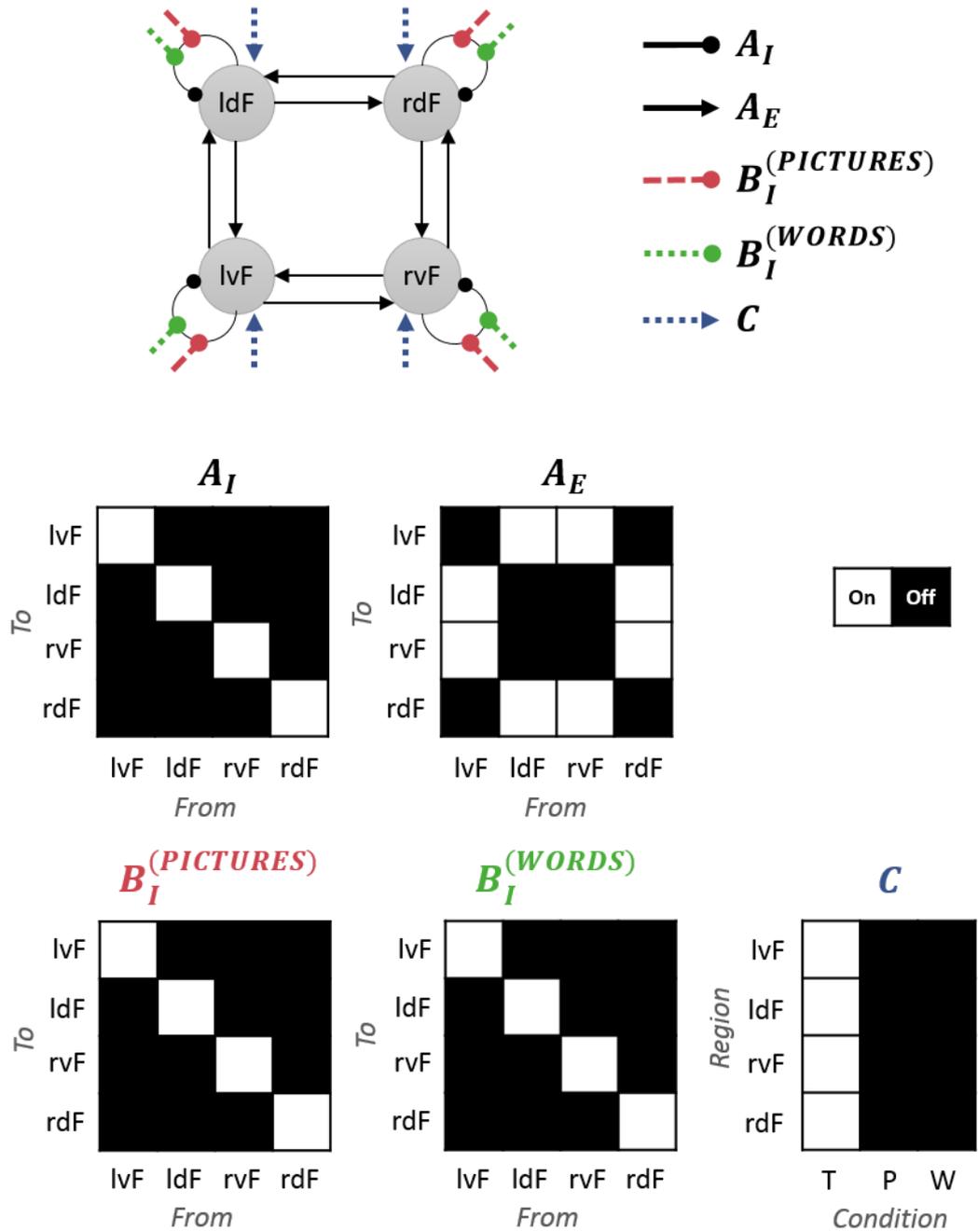

Figure 3 The network architecture implemented for this analysis. Top: Schematic of the network indicating which parameters were switched on. These were the average connections over experimental conditions (intrinsic self-connections $A_I$ and extrinsic between-region connections $A_E$), modulation of self-connections by pictures and / or words ($B_I$) and driving input by Task ($C$ matrix). This is a simplification of the architecture used by Seghier et al. (2011). Bottom: The matrices corresponding to this network, indicating which parameters were estimated from the data (switched on, white) and which were fixed at zero (switched off, black). The regions of frontal cortex were left ventral, lvF, left dorsal, ldF, right ventral, rvF, right dorsal, rdF. The experimental conditions in matrix $C$ were T=task, P=pictures, W=words.

# 6 Haemodynamic model specification

The DCM haemodynamic model predicts the fMRI timeseries one would expect to measure, given neural activity. This does not require specification on a per-experiment basis, so here we just provide a brief summary of the pathway from neural activity to fMRI timeseries. Technical details are given in *Appendix 5: Haemodynamic and BOLD signal model*.

Following experimental stimulation, the temporal evolution of the BOLD signal can be divided into deoxygenated, oxygenated and sustained response phases, each of which can be linked to interactions of neuronal activity, neurovascular coupling, and blood vessel dynamics as summarized in Figure 4. The baseline level of the BOLD signal is determined by the net oxygen extraction exchange between neurons and blood vessels, as well as cerebral blood flow. In response to experimental stimulation, neurons consume oxygen, increasing the ratio of deoxygenated to oxygenated blood. This is reflected by a lag in the BOLD response (the deoxygenated phase). In response to stimulation, neural activity drives astrocytes, releasing a vasodilatory signal (e.g., nitric oxide), which causes an increase in cerebral blood inflow. As a result, the oxygen level, blood volume, and blood outflow are all increased, which is accompanied by a rise in BOLD signal (oxygenated phase) up to a peak five to six seconds after stimulation. In the absence of further stimulation, the activity of neurons return to their resting state, accompanied by a gradual decrease in the BOLD signal (sustained response phase). The dynamic interactions between cerebral blood flow, deoxyhemoglobin and blood volume are captured by the haemodynamic model (**spm_fx_fmri.m**) and the BOLD signal model (**spm_gx_fmri.m**), the parameters of which are estimated on a per-region basis. These parameters are concatenated with those of the neural model and estimated using the fMRI data.

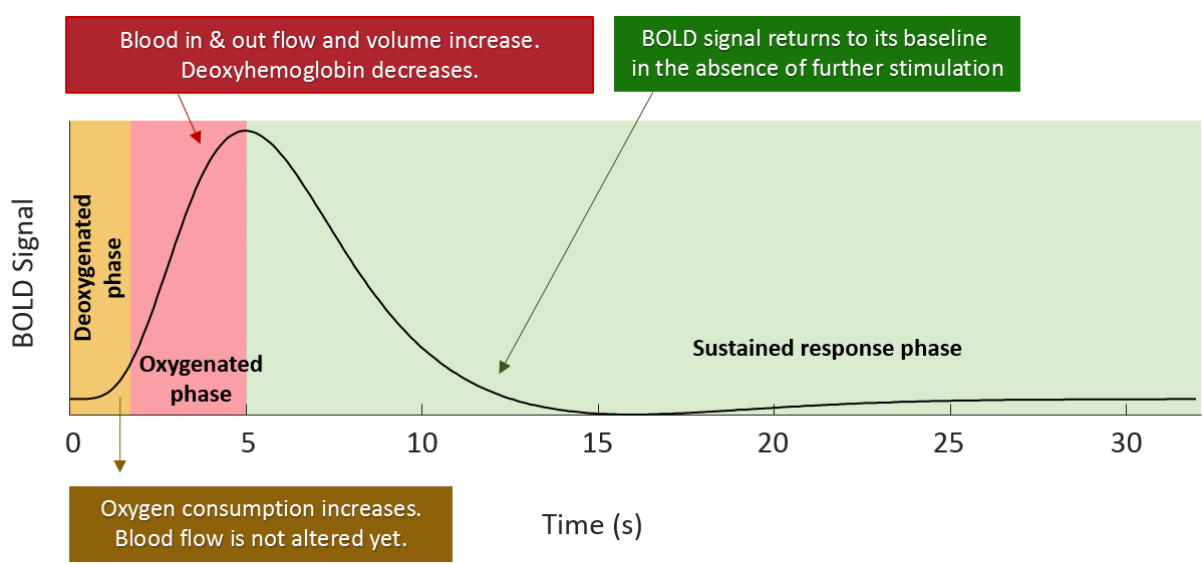

Figure 4 BOLD signal divided into deoxygenated, oxygenated and sustained response phases. The DCM forward model captures the biophysical processes that give rise to this signal. In the deoxygenated phase, neurons consume oxygen while blood flow is not altered. The blood inflow, outflow, and oxygen level increase in response to the neural activity, up to the peak of the BOLD signal at 5-6s post stimulation. BOLD signal exhibits a gradual decay to its baseline in the absence of further stimulation.

# 7 Model estimation

Having specified the forward model, the next step is to invert the model for each subject (**spm_dcm_fit.m**). Estimation or inversion is the process of finding the parameters (e.g. connection strengths) that offer the best trade-off between explaining the data and minimizing complexity (i.e. keeping the parameters close to their prior or starting values). Because there are multiple settings of the parameters that could potentially explain the observed data, DCM uses Bayesian inference, which involves quantifying uncertainty about the parameters before and after seeing the data. This starts with specifying *priors*, which restrict the parameters to a reasonable range. Model estimation combines the priors with the observed fMRI data to furnish updated *posterior* beliefs (i.e. after seeing the data). The priors and posteriors have the form of probability densities. Below, we detail the priors used in DCM, which are configured by the DCM software when model estimation is performed. We will then briefly explain the model estimation procedure itself, known as Variational Laplace.

## 7.1 Priors

The priors over parameters in DCM form a multivariate normal density, which is specified by its mean and covariance. Practically these densities are expressed as a vector of numbers (the mean or expected values of the parameters) and a covariance matrix. Elements on the leading diagonal of the covariance matrix are the prior variance (uncertainty) for each parameter, and the off-diagonal elements are the covariance between the parameters. The choice of priors for each connectivity parameter depends on whether the connection was 'switched on' or 'switched off'. Each switched on parameter has expectation zero and non-zero variance (Figure 5, left). This says that in the absence of evidence to the contrary, we assume there is no connectivity or experimental effect, but we are willing to entertain positive or negative values if the data support it. The width of this distribution (its variance) determines how uncertain we are that the parameter is zero. The prior for each 'switched off' parameter has expectation zero and variance close to zero (Figure 5, right). This says that we are certain that the parameter is zero, regardless of the data. Both of these are called 'shrinkage priors', because, in the absence of evidence, the posterior density shrinks to zero. For this experiment, we selected the connections to switch on and off (Figure 3), and the DCM software translated these choices into priors for each parameter (**spm_dcm_fmri_priors.m**). Note that by default, in order to decrease the time required for model estimation, if more than eight brain regions are included then DCM automatically constrains the model by using functional connectivity based sparsity-inducing priors (Seghier and Friston, 2013). This was not the case here, and the priors for all free parameters are listed in Table 2.

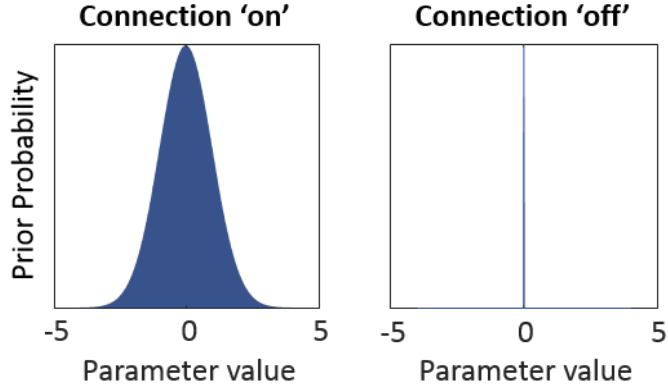

Figure 5 Illustration of priors in DCM. Left: the prior for a 'switched on' parameter is a Gaussian probability density with zero mean and non-zero variance. Right: the prior for a 'switched off' parameter has zero or close-to-zero variance, meaning the parameter is fixed at the prior expectation, which is typically zero.

## 7.2 Variational Laplace

Model inversion (i.e., parameter estimation) is the process of finding the parameters that enable the model to best explain the data; i.e. maximize the log model evidence $\ln p(Y|m)$. This is the log of the probability of having observed the data $Y$ given the model $m$. Generally, model evidence cannot be calculated or derived analytically (because it involves marginalization over very high dimensional integrals); so instead an approximation called the negative variational free energy $F$ (Friston et al., 2007) can be used. The free energy is a lower bound on the model evidence (in machine learning, an Evidence Lower Bound or ELBO). It is useful because it scores how well the model achieved a trade-off between accuracy and complexity:

$$\ln p(Y|m) \cong F = \text{accuracy}(Y,m) - \text{complexity}(m) \qquad (6)$$

The accuracy term quantifies how closely the predicted timeseries corresponds to the observed data. The complexity term is the Kullback-Leibler divergence between the priors and the posteriors; i.e., the difference between the two distributions. If the parameters had to move far from their prior expectation in order to explain the data, then the complexity of the model will be high. This measure of complexity also distinguishes parameters that are independent from those that co-vary (making less individual contributions to explaining the data). When selecting among several models of the same data, the best model is the one with the highest (most positive) free energy, because it offers the most accurate and least complex explanation for the data. We used the DCM software to invert each subject's model, obtaining estimates of their free energy $F$ and the posterior probability density over the parameters that maximised $F$. This completes a description of the first-level (within subject) analysis.

# 8 Results

## 8.1 Diagnostics

A basic diagnostic of the success of model inversion is to look at the estimated parameters and the percentage variance explained by the model. Figure 6 (top) and Table 3 show the neural parameters from a randomly selected subject (subject 37), which we will use to exemplify an interpretation of the parameters (**spm_dcm_review.m**). Many of the neural parameters ($A, B, C$) moved away from their prior expectation of zero, with 90% credible intervals (pink bars) that do not include zero. Figure 6 (bottom) shows the modelled timeseries and residuals from this subject. There were clearly dynamics (solid lines) related to the onsets of the task (grey boxes). The explained variance for this subject was 18.85% and the mean across subjects was 17.27% (SD 9.37%), computed using **spm_dcm_fmri_check.m**. It is unsurprising that the explained variance was quite low, because we did not model the control conditions (perceptual matching) or the baseline rest periods. Nevertheless, most of the subjects evinced nontrivial neural parameters, with 90% confidence intervals that excluded zero; so we could be confident that there was useful information in the data pertaining to our experimental effects.

## 8.2 Interpretation of parameters

We will use the same subject's model to interpret key parameters. The $B$ parameters are the most interesting experimentally; these are the modulations of connections by each experimental condition (*Pictures* and *Words*). Positive parameter estimates indicate increased self-inhibition due to the experimental condition, and negative values meant disinhibition. We allowed picture and word stimuli to modulate each of the self-connections, and three of these parameters, numbered 13, 14 and 17, deviated with a high degree of posterior confidence from their prior expectation of zero. These are plotted in Figure 6 (top) and are illustrated in green and red text in Figure 7. Picture stimuli increased self-inhibition on ldF and decreased self-inhibition on lvF, thereby shifting responses from the dorsal to ventral frontal cortex, specifically in the left hemisphere. Word stimuli increased self-inhibition in lvF, making it less sensitive to input from the other modelled regions.

It is sufficient to report the estimated parameters and make qualitative statements about their meaning, as above (e.g., that the strength of a particular connection was increased or decreased by an experimental condition). However, what is the quantitative interpretation of these parameters? Taking region lvF as an example, we can write out Equation 3 in full, to express the rate of change in lvF's neural activity. This is shown in Equation 7:

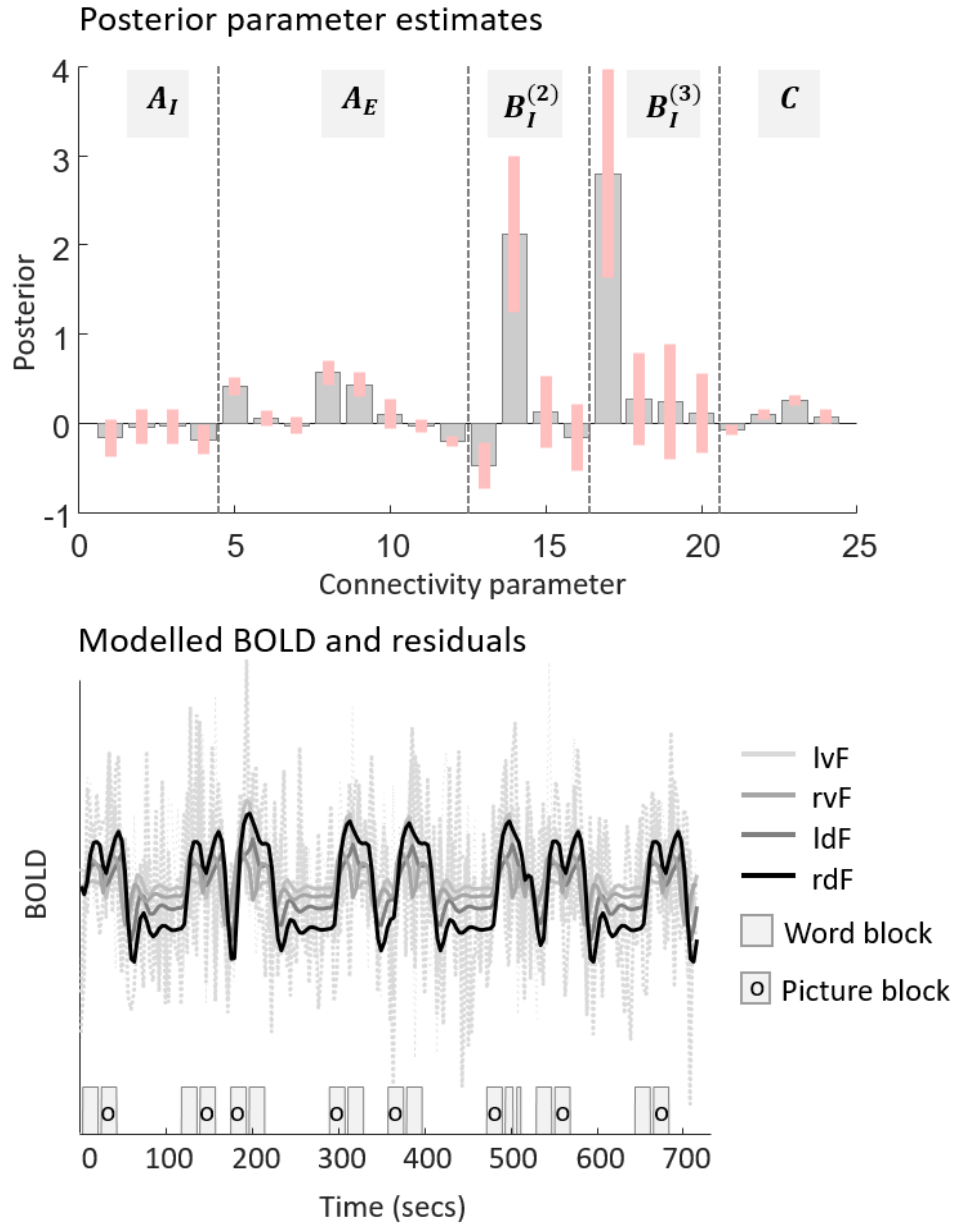

Figure 6 Example DCM neural parameters and model fit for a single subject. **Top**: The parameters corresponding to Equation 3. The error bars are 90% credible intervals, derived from the posterior variance of each parameter, and the vertical dotted lines distinguish different types of parameter. Note this plot does not show the covariance of the parameters, although this is estimated. The parameters are: the average inhibitory self-connections on each region across experimental conditions ($A_I$), the average between-region extrinsic connections ($A_E$), the modulation of inhibitory self-connections by pictures ($B_I^{(2)}$) and by words ($B_I^{(3)}$), and the driving inputs ($C$). For a full list of parameters, please see Table 3. **Bottom**: Example subject's predicted timeseries (solid lines) with one line per brain region. The dotted lines show the model plus residuals. Underneath, blocks showing the timing of the word and picture trials.

$$\dot{z}_1 = \underbrace{\left( \underbrace{-0.5 \cdot \exp(A_{I\,11})}_{\text{Average}} \cdot \underbrace{\exp\left(B^{(2)}_{I\,11} \cdot u_2(t)\right)}_{\text{Pictures}} \cdot \underbrace{\exp\left(B^{(3)}_{I\,11} \cdot u_3(t)\right)}_{\text{Words}} \right) z_1}_{\text{Self-connection}}$$

$$+ \underbrace{A_{E\,12} \cdot z_2}_{\text{ldF}\rightarrow\text{lvF (A)}} + \underbrace{A_{E\,13} \cdot z_3}_{\text{rvF}\rightarrow\text{lvF (A)}} + \underbrace{C_{11} \cdot u_1(t)}_{\text{Driving (C)}}$$

$$u_1(t) = \begin{cases} 0.6, & task \\ -0.4, & otherwise \end{cases}$$

$$u_2(t) = \begin{cases} 0.8, & pictures \\ -0.2, & otherwise \end{cases}$$

$$u_3(t) = \begin{cases} 0.8, & words \\ -0.2, & otherwise \end{cases}$$

(7)

This says that the response in region lvF was governed by the strength of its self-connection (line 1 of Equation 7) as well as incoming connections from regions ldF, rvF and the driving input (line 2 of Equation 7). The values for the experimental inputs $u_1(t)$, $u_2(t)$ and $u_3(t)$ at time $t$ were set during the specification of the model, due to mean-centring of the regressors (see Section 5.6: Centre input). Plugging in the estimated parameters from Table 3, the self-inhibition in lvF during picture trials was $-0.5 \cdot \exp(-0.16) \cdot \exp(-0.47 \cdot 0.8) \cdot \exp(2.8 \cdot -0.2) = -0.17 Hz$. The self-inhibition of lvF during word trials was far stronger: $-0.5 \cdot \exp(-0.16) \cdot \exp(-0.47 \cdot -0.2) \cdot \exp(2.8 \cdot 0.8) = -4.40 Hz$. Therefore, region lvF was more sensitive to inputs from the rest of the network when the stimuli were pictures than words. These task effects can also be expressed as a change in the time constant $\tau$ of region lvF: $\tau = 5.88$s in the context of pictures and $\tau = 0.23$s in the context of words (see Appendix 4). Rewriting this as the half-life of region lvF; neural activity decayed to half its starting level 4.08s after the onset of picture stimuli and 0.16s after the onset of word stimuli. Picture stimuli therefore elicited a far more sustained response in lvF than word stimuli. The other key factor influencing lvF was the incoming connection from region rvF (0.43Hz), and the positive sign indicates this connection was excitatory.

Inspecting the parameters in this way provides insight into the sign and magnitude of the connection strengths and experimental effects. However, this does not constitute a formal test of any hypotheses. There are various strategies for testing hypotheses at the group (between-subject) level, using classical or Bayesian statistics, and we detail these in the second part of the tutorial (please see the companion paper).

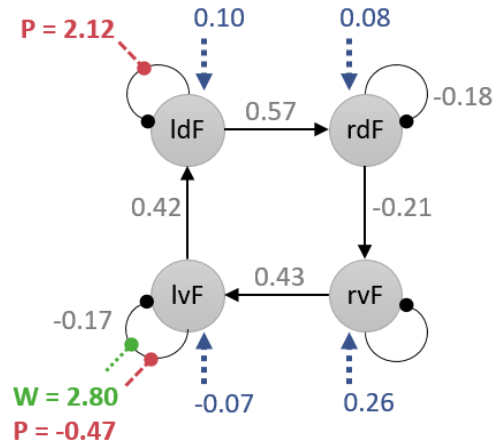

Figure 7 Estimated parameters from a single subject. Between-region (extrinsic) parameters are in units of Hz, where positive numbers indicate excitation and negative numbers indicate inhibition. Self-connection parameters have no units and scale up or down the default self-connection of -0.5Hz (see Equation 3). Positive numbers for the self-connections indicate increased self-inhibition and negative numbers indicate disinhibition. For clarity, only parameters with 90% probability of being non-zero are displayed (see Table 3 for details). Colours and line styles as for Figure 3.

# 9 Discussion

This tutorial reviews the current implementation of DCM for fMRI by stepping through the analysis of a factorial fMRI experiment. This first level (within-subject) analysis started by identifying brain regions evincing experimental effects, for which we extracted representative fMRI timeseries. We then specified a DCM, by selecting which connections should be 'switched on' and which should be 'switched off'. This specified the priors for the connectivity parameters. Inverting each subject's model provided the strength of connections ($A$), the change in connections due to each experimental condition ($B$) and the sensitivity of each region to external input ($C$), as well as the free energy approximation to the log model evidence $F$. The appendices provide the technical detail of each of these steps.

A common question from users is: what assumptions are made by DCM? As a Bayesian model, most assumptions are stated up-front as priors. The key assumptions for the basic (deterministic 1-state bilinear) neural model are as follows:

- The pre-processed fMRI timeseries used for DCM have been selected because they show experimental effects. The signals are averaged over voxels and nuisance effects are regressed out, therefore, the signal-to-noise ratio is high – the prior expectation of the variance of the noise is $\frac{1}{\exp(6)} = 0.0025$ (see Appendix 2). This expresses the prior belief that most of the variance is interesting and where possible, we would like the variance to be ascribed to the model rather than to observation noise. Furthermore, the variance of the observation noise is assumed to be independent of the neural / haemodynamic parameters.

- The neural response due to intrinsic (within-region) activity is expected to decay over a period of seconds following experimental stimulation. The prior on the self-connection parameters says that an isolated brain region's time constant τ will be between 1.63s and 2.46s with 90% probability, and between 0.38s and 10.49s in the context of modulation by an experimental condition (Appendix 4). This response will be further added to or subtracted from by incoming connections from other regions.

- The priors for the parameters of the haemodynamic, BOLD signal and observation models are consistent with empirical measurements using animal models and human subjects (c.f. Buxton et al., 1998; Stephan et al., 2007). In DCM for fMRI, three of these parameters are estimated from the data and the priors are listed in Table 2. Values for fixed parameters, which are not estimated from the data, can be found in the Matlab functions **spm_fx_fmri.m** and **spm_gx_fmri.m**.

- The free energy is assumed to serve as a good proxy for the log model evidence. This is exactly true for linear models (where the free energy becomes log model evidence) and has been validated for weakly non-linear models like DCM for fMRI using sampling methods (Chumbley et al., 2007). Caution needs to be taken with highly nonlinear models, where local optima pose a challenge; one method for addressing this is to use a multi-start estimation algorithm which re-initializes subject-level inversions using group-level estimated parameters (Friston et al., 2015).

The next step in our analysis was to test which neural effects were conserved over subjects, and which differed due to brain Laterality Index – the between-subjects factor that was the focus of this experiment. These analyses are detailed in the companion paper, where we cover Bayesian model comparison (i.e., hypothesis testing) at the within and between subject level.

# 10 Appendix 1: Timeseries extraction

Before a DCM can be specified, Regions of Interest (ROIs) need to be selected and representative timeseries extracted from each. The fMRI data for a subject can be considered a large 4D matrix $\hat{Y}$ where the first three dimensions are space and the fourth dimension is time (in scans). By extracting timeseries, we seek to reduce this to a smaller matrix $Y$ where there are a small number of ROIs that define our brain network. There are various strategies for selecting the voxels that contribute to each ROI – indeed, questions pertaining to this are among the most common from DCM users on the SPM Mailing List. The most important consideration is that DCM is intended to explain the coupling between neural populations *that show experimental effects*. An initial GLM analysis is therefore normally used to identify voxels that show a response to each experimental factor. To reduce noise, only voxels that exceed some liberal statistical threshold for a contrast of interest are usually retained.

For the data presented here, the following steps were applied by (Seghier et al., 2011), which may provide a useful recipe for preparing DCM studies:

1. **Statistical Parametric Mapping (SPM).** A General Linear Model (GLM) was specified for each subject, and T-contrasts were computed to identify brain regions that showed a main effect of each factor and an interaction between factors. Additionally, an F-contrast was calculated to identify all 'Effects of Interest' – to later regress out any uninteresting effects such as head motion or breathing from the timeseries. This F-contrast was an identity matrix of dimension $n$, where the first $n$ columns in the design matrix related to interesting experimental effects.

2. **Group-level region selection**. Contrast maps from each subject were summarized at the group level using one-sample t-tests. These group-level results were used to select the peak MNI coordinates of the ROIs. Different contrasts could have been used to select each ROI; however, in this case, the main effect of task (semantic > perceptual matching) was used to identify all four ROIs.

3. **Subject-level feature selection**. Having identified the ROI peak coordinates at the group level, the closest peak coordinates for each individual subject were identified. This allowed for each subject to have slightly different loci of responses. Typically, one would constrain each subject-level peak to be within a certain radius of the group-level peak, or alternatively, to be within the same anatomical structure (e.g. using an anatomical mask). Here, subject-level peaks were constrained to be a maximum of 8mm from the group level peak, and had to exceed a liberal statistical threshold of $p < 0.05$ uncorrected.

At this stage, Seghier et al. (2011) excluded any subjects not showing experimental effects in every brain region above the statistical threshold. We suggest that with the development of hierarchical modelling of connectivity parameters, detailed in the second part of this tutorial, removing subjects with noisy or missing data in certain brain regions may be unnecessary. A subject who lacks a strong response in one brain region or experimental condition, for whatever reason, may still contribute useful information about other brain regions or conditions (and indeed useful information about intersubject variability). Therefore, when an ROI contains no voxels showing a response above the selected threshold, we recommend dropping the threshold until a peak voxel coordinate can be identified.

4. **ROI definition**. Having identified the peak coordinates for each ROI, timeseries were extracted. Each ROI was defined as including all the voxels which met two criteria: 1) located within a sphere centred on the individual subject's peak with 4mm radius and 2) exceeded a threshold of $p < 0.05$ uncorrected, for the task contrast at the single-subject level. Note that applying a threshold at this stage is not to ensure statistical significance (this happens in step 2). Rather, the threshold is simply used to exclude the noisiest voxels from the analysis.

5. **ROI extraction**. SPM was used to extract representative timeseries from each ROI, which invoked a standard series of processing steps (**spm_regions.m**). The timeseries are pre-whitened (to reduce serial correlations), high-pass filtered, and any nuisance effects not covered by the Effects of Interest F-contrast are regressed out of the timeseries (i.e. 'adjusted' to the F-contrast). Finally, a single representative timeseries is computed for each ROI by performing a principal components analysis (PCA) across voxels and retaining the first component (or principal eigenvariate). This approach is used rather than taking the mean of the timeseries, because calculating the mean would cause positive and negative responses to cancel out (and further that means are effected by extreme values). That could pose a problem due to centre-surround coding in the brain, where excitatory responses are surrounded by inhibitory responses – and would cancel if averaged.

Additionally, prior to DCM model estimation, the software automatically checks whether the fMRI data are within the expected range (**spm_dcm_estimate.m**). If the range of the fMRI data exceeds four (in the units of the data), DCM rescales the data to have a range of four, on a per-subject basis. This was the case for our data.

These steps produced one timeseries per region, for each subject, which were then entered into the DCM analysis. The complete pipeline above can be performed in the SPM software semi-automatically, using the steps described in the practical guide.

# 11 Appendix 2: Observation noise specification

DCM separately estimates the precision (inverse variance) of zero-mean additive white noise for each brain region (**spm_nlsi_gn.m**). The white noise assumption is used because the preliminary general linear model estimates serial correlations, which are used to whiten principal eigenvariates from each region. From Equation 1 we have the model:

$$Y = g(z, \theta^{(h)}) + X_0 B_0 + \epsilon \qquad (8)$$

To simplify the implementation, $Y$ is vectorised (the timeseries from each region are stacked on top of one another) to give $y_v = vec(Y)$. The observation noise $\epsilon$ is specified according to a normal density:

$$\epsilon \sim N(0, \Sigma_y) \qquad (9)$$

In practice, DCM uses the precision matrix $\Pi_y$ which is the inverse of the covariance matrix $\Sigma_y$. It is specified by a multi-component model, which is a linear mixture of precision components $Q_i$ with one component per brain region $i = 1 \ldots R$. Each precision matrix is weighted by a parameter $\lambda_i$ which is estimated from the data:

$$\mathbf{\Pi}_y = \sum_i \exp(\lambda_i)\, \mathbf{Q}_i \qquad (10)$$

The diagonal elements of the precision matrix $\mathbf{Q}_i$ have value one for observations associated with brain region $i$ and zero elsewhere. Taking the exponential of parameter $\lambda_i$ ensures that the estimated precision cannot be negative. In total, in the experiment presented here, we had 792 observations per subject ($T = 198$ fMRI volumes times $R = 4$ brain regions), and the corresponding precision components are illustrated in Figure A.1.

From Table 2, the prior density for parameter $\lambda_i$ was $N\left(6, \frac{1}{128}\right)$. This means that scaling factor $\exp(\lambda_i)$ had a lognormal prior density: $Lognormal\left(6, \frac{1}{128}\right)$. The resulting prior expected precision was $\exp(6) = 403.43$ with 90% credible interval [348.84 466.56]. This prior says that the data were expected to have a high signal-to-noise ratio, because the fMRI data were highly pre-processed and averaged, and are selected from brain regions that are known to show experimental effects. Model inversions are therefore preferred which ascribe a high level of variance to the model rather than to noise.

$$\mathbf{\Pi}_y = \lambda_1 \underbrace{Q_1\,(lvF)}_{T \cdot R} + \lambda_2\, Q_2\,(ldF) + \lambda_3\, Q_3\,(rvF) + \lambda_4\, Q_4\,(rdF)$$

Figure A.1 Illustration of the observation noise model in DCM for fMRI. Each precision component $\mathbf{Q}$ was a matrix with $T \cdot R = 792$ elements, and there was one precision component per brain region. Log scaling parameter $\lambda_i$ was estimated from the data and scaled up or down the corresponding component $\mathbf{Q}_i$.

# 12 Appendix 3: Derivation of the fMRI neural model

Neural responses may be written generically as follows:

$$\dot{\mathbf{z}} = \frac{d\mathbf{z}}{dt} = f(\mathbf{z}, \mathbf{u}) \qquad (11)$$

Where vector $\mathbf{z}$ is the state or level of activity in each region, $\dot{\mathbf{z}}$ is the rate of change in each brain region – called the neural response - and $f$ is a function describing the change in brain activity in response to experimental inputs $\mathbf{u}$. The 'true' function $f$ would be tremendously complicated, involving the nonlinear, complex and high dimensional dynamics of all cell types involved in generating a neural response. Instead, we can approximate $f$ using a simple mathematical tool – a Taylor series. The more terms we include in this series, the closer we get to reproducing the true neural response. The definition of the Taylor series $T$ up to the second term, with two variables $z$ and $u$ evaluated at $z = m$ and $u = n$ is:

$$\begin{aligned}
T(z,u) = {} & f(m,n) \\
& +(z-m)\cdot f_z \\
& +(u-n)\cdot f_u \\
& +\frac{1}{2}((z-m)^2\cdot f_{zz} + 2(z-m)(u-n)\cdot f_{zu} + (u-n)^2\cdot f_{uu})
\end{aligned} \qquad (12)$$

Where $f_z$ and $f_u$ are the partial derivatives of $f$ with respect to $z$ and $u$, $f_{zz}$ and $f_{uu}$ are the second order derivatives and $f_{zu}$ is the mixed derivative (i.e. the derivative of $f$ with respect to $z$ of the derivative with respect to $u$, or vice versa). Each of these partial derivatives is evaluated at $(z=m, u=n)$. Setting $m=0$ and $n=0$, defining the baseline neural response $f(m,n)=0$ and dropping the higher order terms we get the simpler expression:

$$\begin{aligned}
T(z,u) = {} & 0 \\
& +z\cdot f_z \\
& +u\cdot f_u \\
& +\frac{1}{2}(z^2\cdot f_{zz} + 2zu\cdot f_{zu} + u^2\cdot f_{uu})
\end{aligned} \qquad (13)$$

By re-arrangement of the final term:

$$\begin{aligned}
T(z,u) = {} & z\cdot f_z \\
& +u\cdot f_u \\
& +zu\cdot f_{zu} \\
& +\frac{1}{2}(z^2\cdot f_{zz} + u^2\cdot f_{uu})
\end{aligned} \qquad (14)$$

Finally, factorizing $z$ and dropping the final term (as $z^2$ and $u^2$ will be very small around the origin) gives:

$$\begin{aligned}
T(z,u) &= (f_z + u\cdot f_{zu})z + u\cdot f_u \\
&= (A+Bu)z + Cu
\end{aligned} \qquad (15)$$

Here, we have assigned letters to the three derivative terms $A=f_z, B=f_{zu}, C=f_u$, which gives the expression for the neural model used in in the DCM literature (Equation 2). With multiple brain regions, these becomes matrices. As introduced in the main text, **A** is the rate of change in neural response due to the other neural responses in the system – i.e. the effective connectivity. **B** is the rate of change in effective connectivity due to the inputs and is referred to as the bilinear or interaction term. Finally, **C** is the rate of change in neural response due to the external input, referred to as the driving input. In the DCM framework, **A**, **B** and **C** become parameters which are estimated from the data. (To apply negativity constraints on the self-connections, **A** and **B** are sub-divided into intrinsic and extrinsic parts, see Equation 3.)

# 13 Appendix 4: The neural parameters

Whereas Appendix 3 motivated the DCM neural model as a function approximated by a Taylor series, here we consider it from the perspective of a simple dynamical system, to help gain an intuition for the parameters. Consider a DCM with a single brain region, driven by a brief stimulus at time $t = 0$. The neural equation can be simplified to the following:

$$\dot{z} = az \quad (16)$$

Where self-connection or rate constant $a$ has units of Hz and is negative. The solution to this equation, the neural activity at any given time, is an exponential decay (under the constraint that $a$ is negative):

$$z(t) = z(0) \cdot \exp(at) \quad (17)$$

Where $z(0)$ is the initial neuronal activity. This function is plotted in Figure A.2 (left) with parameter = $-0.5 Hz$, which is the default value in DCM.

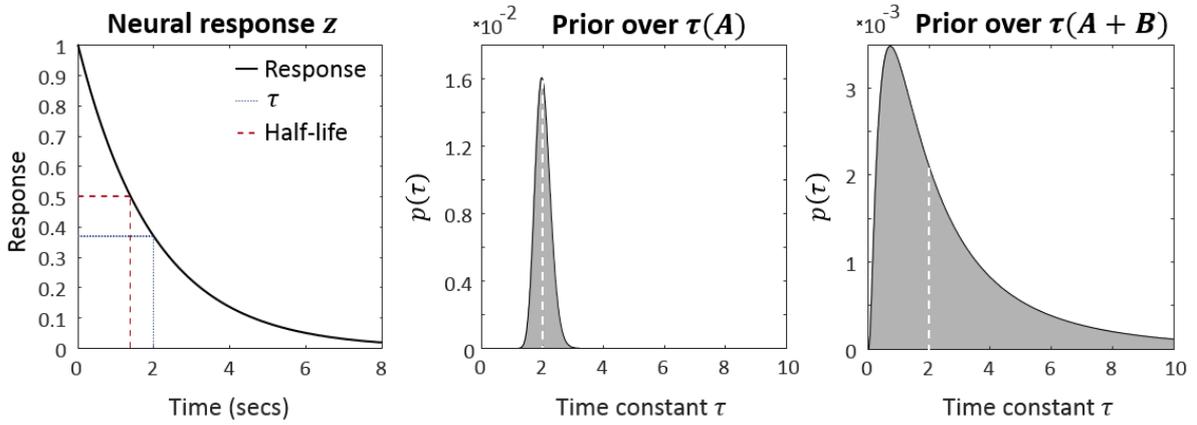

Figure A.2 Illustration of neural response as an exponential decay. Left: the neural response under the default prior of a = -0.5Hz to an instantaneous input at time zero. Also plotted are the corresponding time constant $\tau$ and half-life. Middle: The resulting prior over time constant $\tau$ in the absence of modulation. The median is $\tau = 2$ seconds with 90% credible interval [1.63s 2.46s]. Right: The prior over $\tau$ in the presence of modulation. The median is $\tau = 2$ seconds with 90% credible interval [0.38s 10.49s]. Green dashed lines in the middle and right panels show the median.

Figure A.2 (left) also illustrates two common ways of characterizing the rate of decay. The *time constant $\tau$* is defined as:

$$\tau = -\frac{1}{a} \quad (18)$$

This is the time in seconds taken for the neural activity to decay by a factor of $\frac{1}{e}$ (36.8% of its peak response). Given $a = -0.5Hz$ the time constant is $\tau = 2s$. This inverse relationship between the rate constant $a$ and time constant $\tau$ is why the connectivity parameters in DCM are in units of $Hz$ ($Hz$ is 1/seconds). It can be more intuitive to express the rate of decay as the half-life, which is the time at which the activity decays to half its starting value. Given the self-connection of -0.5Hz, the half-life is:

$$t_{\frac{1}{2}} = \tau \cdot \ln 2 = 1.39\text{s} \tag{19}$$

In DCM we specify a prior probability density over each self-connection parameter $a$, which in turn specifies our expectation about a typical region's time constant. Figure A.2 (centre) shows the resulting prior time constant in DCM for fMRI. The median is 2s with 90% of the probability mass (the credible interval) between 1.63s and 2.46s.

In our analyses, we allowed self-connections to be modulated by experimental conditions. The rate constant $a$ was therefore supplemented to give $a + b \cdot u$ (see Equation 2 of the main text). The modulatory parameter $b$, multiplied by the experimental input $u$, could increase or decrease the region's rate of decay. Figure A.2 (right) shows the prior time constant for connections with modulation switched on (where $u = 1$), giving 90% credible interval [0.38s 10.49s]. These plots make clear that DCM for fMRI does not model the activity of individual neurons, which typically have time constants on the order of milliseconds. Rather, it models the slow emergent dynamics that evolve over seconds and arise from the interaction of populations of neurons. For details of how these plots were generated, please see the supplementary text.

# 14 Appendix 5: Haemodynamic and BOLD signal model

The translation of neural activity $z$, predicted by the DCM neural model, to observed BOLD response $y$, is described by a three-part model illustrated in Figure A.2. We will summarise each of the three parts in turn.

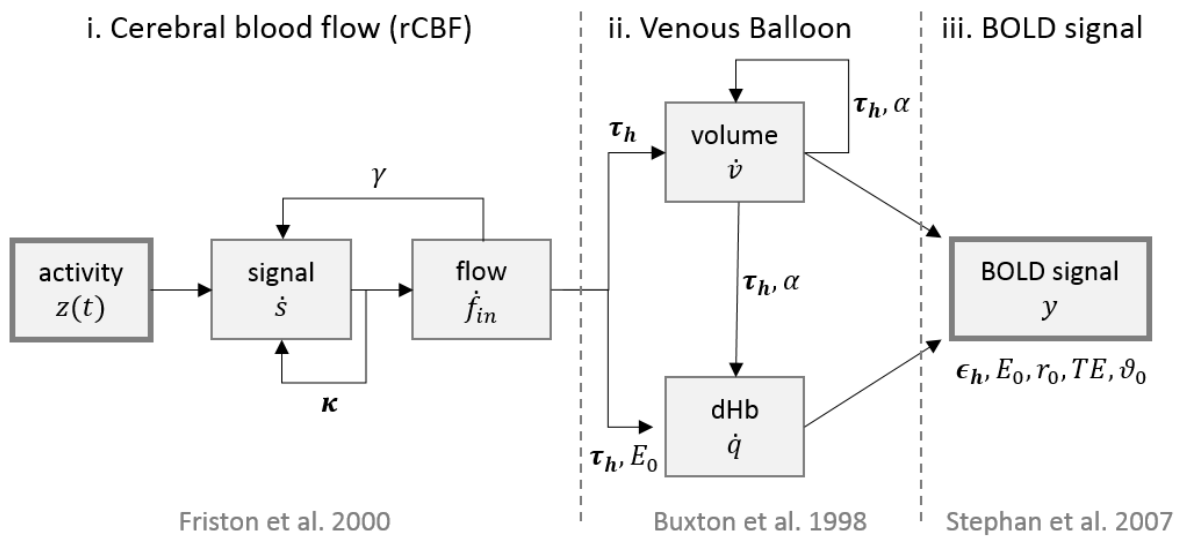

Figure A.2 The model used to translate from neural activity to the BOLD signal in DCM. This is split into three parts. **i.** Neural activity $z(t)$ triggers a vasoactive signal $s$ (such as nitric oxide) which in turn causes an increase in blood flow. **ii.** The flow inflates the blood vessel like a balloon, causing a change in both blood volume $v$ and deoxyhaemoglobin (dHB) $q$. **iii.** These combine non-linearly to give rise to the observed BOLD signal. A key reference for each part of the model is given - see text for further details. Symbols outside the boxes are parameters and those in bold type are free parameters that are estimated from the data: decay $\kappa$, transit time $\tau_h$ and ratio of intravascular to extravascular contribution to the signal $\epsilon_h$. See Table 1 for a full list of symbols. Adapted from Friston et al. (2000).

## 14.1 Regional cerebral blood flow (rCBF)

Neural activity $z(t)$ drives the production of a chemical signal $s$ that dilates local blood vessels. This causes oxygenated blood to flow into the capillaries, where oxygen is extracted. As a result, partially deoxygenated blood flows into the veins (venous compartment). The vasodilatory signal $s$ decays exponentially and is subject to feedback by the blood flow $f_{in}$ that it induces:

$$\dot{f}_{in} = s$$
$$\dot{s} = z(t) - \kappa s - \gamma(f_{in} - 1) \quad (20)$$

Where parameter $\kappa$ is the rate of decay for the signal $s$, and $\gamma$ is the time constant controlling the feedback from blood flow. Empirical estimates have shown $\kappa$ to have a half-life of around one second (Friston et al., 2000), placing it in the correct range to be mediated by nitric oxide (NO). Adjusting $\kappa$ primarily changes the peak height of the modelled BOLD response, whereas adjusting $\gamma$ primarily changes the duration of response. Both parameters also modulate the size of the post-stimulus undershoot.

## 14.2 Venous balloon

Increased blood flow causes a local change in the volume of blood $v$ in the blood vessel, as well as the proportion of deoxyhaemoglobin $q$ (dHb). This process is captured by the Balloon model of Buxton et al. (1998). It treats the venous compartment as a balloon, which inflates due to increased blood flow and consequently expels deoxygenated blood at a greater rate. The change in blood volume $v$, normalized to the value at rest, depends on the difference blood inflow and outflow:

$$\tau_h \dot{v} = f_{in}(t) - f_{out}(v, t)$$
$$f_{out}(v, t) = v(t)^{\frac{1}{\alpha}} \quad (21)$$

Where the time constant $\tau_h$ is the mean transit time of blood, i.e. the average time it takes for blood to traverse the venous compartment. Grubb's parameter $\alpha$ controls the stiffness of the blood vessel (Grubb et al., 1974) and adjusting it has the effect of changing the peak height of the modelled BOLD response.

The increase in blood volume following neural activity is accompanied by an overall decrease in dHb, the rate of which depends on the delivery of dHb into the venous compartment minus the amount expelled:

$$\tau_h \dot{q} = f_{in}(t) \frac{1 - (1 - E_0)^{\frac{1}{f_{in}}}}{E_0} - \frac{f_{out}(v, t)(q(t))}{v(t)} \quad (22)$$

The first expression on the right hand side of Equation 22 approximates the fraction of oxygen extracted from the inflowing blood, which depends on the inflow $f_{in}$ and the resting oxygen extraction fraction $E_0$ (the percentage of the oxygen removed from the blood by tissue during its passage through the capillary network). The second term relates to the outflow, where the ratio $q/v$ is the dHb concentration.

## 14.3 BOLD signal

Finally, the change in blood volume $v$ and dHb $q$ combine to cause the BOLD signal $S$, measured using fMRI. For the purpose of this paper we define $S$ as the signal acquired using a gradient echo EPI readout. The model used in DCM is due to Buxton et al. (1998) and Obata et al. (2004), which were extended and re-parameterised by Stephan et al. (2007). In the following paragraphs we provide a recap of the basic mechanisms of MRI and functional MRI (fMRI), in order to motivate the form of the BOLD signal model. Readers familiar with MR physics may wish to skip this introduction.

We will use classical mechanics to describe the way the MR signal is generated. When entering an MRI scanner, the subject is exposed to the main magnetic field $\boldsymbol{b_0}$. This magnetic field is always on and its axis is aligned with the tunnel of the scanner. All the hydrogen protons of the body can be thought of as acting like tiny magnets whose strength is measured by their magnetic moments $\boldsymbol{\mu}$. In what follows, the coordinate system x,y,z is used where z corresponds to the $\boldsymbol{b_0}$ axis, y and x are the orthogonal vectors forming the transverse plane. When submitted to the magnetic field $B_0$, two phenomena occur:

1/ All of the proton magnetic moments precess about the $\boldsymbol{b_0}$ axis at the Larmor frequency, which is proportional to the amplitude of the $\boldsymbol{b_0}$ field strength (e.g. 123MHz at 3T).

2/ The proton magnetic moments orient themselves such that their vector sum is a net magnetization vector, $\boldsymbol{m}$, aligned with the $\boldsymbol{b_0}$ field axis and pointing in the same direction (Figure A.3.i). The net magnetization vector $\boldsymbol{m}$ can be decomposed into two components, the longitudinal component $\boldsymbol{m_z}$ along the z axis and the transverse component $\boldsymbol{m_{xy}}$, which is the projection of $\boldsymbol{m}$ into the transverse plane. The transverse component is zero when the system is at equilibrium; i.e., when the net magnetization is aligned with the z axis, yet it is only the transverse component that can be measured in MRI.

In order to disturb the equilibrium state and thereby create a transverse component $\boldsymbol{m_{xy}}$ that can be detected, a rotating magnetic field $\boldsymbol{b_1}$ is applied orthogonal to the $\boldsymbol{b_0}$ axis for a short period of time. This is termed 'excitation' and results in the tilting of the net magnetization towards the transverse plane (Figure A.3.ii).

Once the $\boldsymbol{b_1}$ field is turned off, the net magnetization has a transverse component and continues to precess around the main magnetic field, $\boldsymbol{b_0}$. Since the precession frequency is proportional to the amplitude of the magnetic field, any spatial variation of the magnetic field amplitude across one voxel will induce a difference in precessional frequency for the protons. For this reason, the protons accumulate a delay relative to each other and so have differential phase (orientation, Figure A.3.iii). Over time the delays, or relative phase difference, increase (Figure A.3.iv). As a result their vector sum; i.e., the transverse component $\boldsymbol{m_{xy}}$ decreases. This process, whereby the transverse component of the net magnetisation decreases, is called effective transverse relaxation. It is characterized by an exponential decay with a time constant $T_2^*$ (Figure A.3.v), or alternatively a relaxation rate $R_2^*$, whereby $R_2^* = \frac{1}{T_2^*}$.

Crucially, for functional MRI, dhB and oxyhaemoglobin (Hb) molecules have different magnetic susceptibility (i.e. a different response to being placed in a magnetic field). Unlike Hb, which exhibits a weak, diamagnetic response to the main magnetic field, dHb exhibits a stronger, paramagnetic response. At the boundaries between two tissues with different magnetic susceptibilities, the magnetic field is distorted, increasing the local spatial inhomogeneity in the amplitude of the magnetic field. The decrease in dHb following neural activity makes the blood less paramagnetic, and more similar to the surrounding tissue in terms of magnetic susceptibility. As a result, the magnetic field around the blood vessel becomes less distorted, with a smaller range of precessional frequencies of protons in the voxel. As a consequence, less differential phase accumulates between the proton magnetic moments and the amplitude of the transverse component of the net magnetization vector $\boldsymbol{m_{xy}}$ decreases less rapidly. This corresponds to a shorter $R_2^*$ (or equivalently a longer $T_2^*$). Therefore, at the time the data are acquired, TE (Echo Time), the signal will be higher if it follows a period of neural activity.

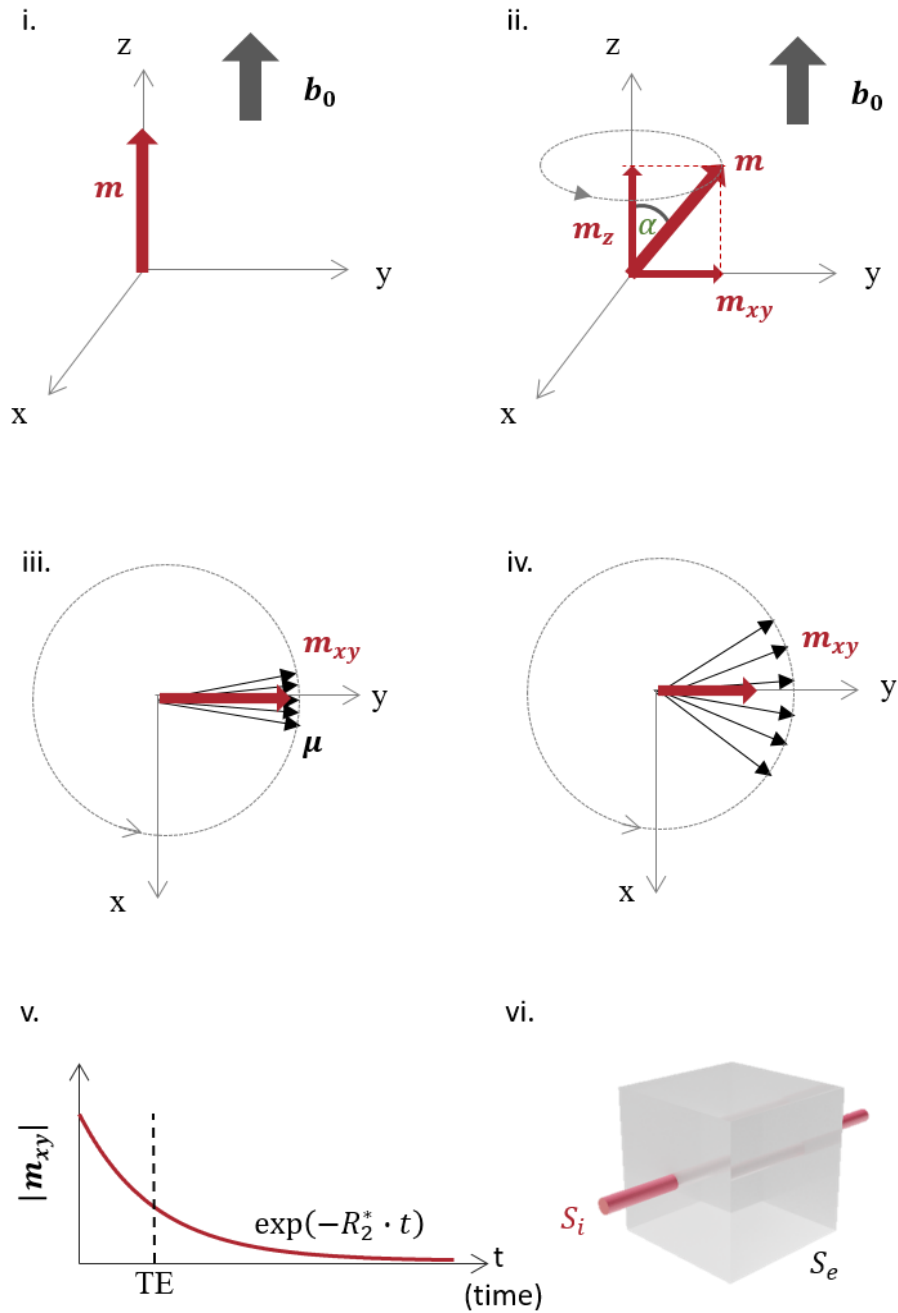

**Figure A.3 Generation of the Magnetic Resonance (MR) signal. i**. When exposed to a strong magnetic field, proton magnetic moments **μ** add together to create a net magnetization **m**, aligned

with the B0 axis pointing in the same direction as this main field. **ii** After excitation with a flip angle ($\alpha$) imparted by a rotating B1 field applied orthogonal to B0 for a short period of time, the net magnetization is composed of a longitudinal and a transverse component, which precesses about the B0 axis. **iii** Inhomogeneity in the magnetic field within a voxel causes the protons' magnetic moments to precess with different frequency leading to differential phase (orientation) between the protons, reducing the transverse component **m**$_{xy}$ which is the vector sum of all the protons. **iv**. The differential phase accumulated by the protons increases over time. **v**. As a result, the amplitude of the transverse component of the net magnetization (i.e. the detectable MR signal) further decreases, following an exponential decay characterized by the effective transverse relaxation rate R2*=1/T2*. The MR signal is acquired at an echo time TE. **vi**. DCM assumes that there are two contributions to the measured signal – intravascular ($S_i$) and extravascular ($S_e$) – each with their own $R_2^*$ relaxation rates.

Having revised the fundamentals of fMRI, we now return to the BOLD signal model in DCM. It follows Ogawa et al. (1993), in treating the tissue within a voxel as consisting of many small cubes, each with a cylinder running through the centre (Figure A.3vi). There are two compartments – the extravascular tissue outside the cylinder and the blood vessel (intravascular venous compartment) that is filled with blood. The BOLD signal at rest $S_0$ is modelled as a linear mixture of these extravascular and intravascular contributions (Buxton et al., 1998):

$$S_0 = (1 - V_0)S_e + V_0 S_i \tag{23}$$

Where $V_0$ is the fraction of the BOLD signal originating from the intravascular compartment. From Obata et al. (2004), each compartment's resting BOLD signal at the time of measurement, $TE$ in seconds, is modelled by:

$$S_e = S_{e0} \cdot \exp(-R_{2e}^* \cdot TE)$$
$$S_i = S_{i0} \cdot \exp(-R_{2i}^* \cdot TE) \tag{24}$$

Where $R_{2e}^*$ and $R_{2i}^*$ are the effective transverse relaxation rates for the extravascular and intravascular compartments respectively, in units of Hz, and $S_{e0}$ and $S_{i0}$ are the maximal signals originating from each compartment before any signal decrease due to differential dephasing of the protons. Following neural activation, there will be an altered BOLD signal, $S$, compared to $S_0$, written $\Delta S = S - S_0$. A linear approximation of this change is as follows:

$$\frac{\Delta S}{S_0} \approx -(\Delta R_{2e}^* \cdot TE) - (V_0 \cdot \epsilon_h \cdot \Delta R_{2i}^* \cdot TE) + (V_0 - V_1)(1 - \epsilon_h) \tag{25}$$

Where $\Delta R_{2e}^*$ and $\Delta R_{2i}^*$ is the change in each compartment's $R_2^*$ between activation and rest, $\epsilon_h = \frac{S_i}{S_e}$ is the ratio of intra- to extra-vascular signal contributions and $V_1$ is the volume of venous blood following activation (Obata et al., 2004; Stephan et al., 2007). Obata et al. (2004) derived approximations for the changes in extravascular and intravascular signal decay:

$$\Delta R_{2e}^* = 4.3 \cdot \vartheta_0 \cdot V_0 \cdot E_0 \cdot (q - 1)$$
$$\Delta R_{2i}^* = r_0 \cdot E_0 \cdot \left(\frac{q}{v} - 1\right) \tag{26}$$

Where $\vartheta_0$ is the frequency offset at the outer surface of magnetised vessels in units of Hz, $E_0$ is the fraction of oxygen extraction at rest, and $r_0$ is a constant relating $\Delta R_{2i}^*$ to the oxygen extraction rate. Plugging these into Equation 25 and re-arranging gives the final expression (Stephan et al., 2007):

$$\frac{\Delta S}{S_0} \approx V_0 \left( k_1(1 - q) + k_2 \left(1 - \frac{q}{v}\right) + k_3(1 - v) \right)$$

$$k_1 = 4.3 \cdot \vartheta_0 \cdot E_0 \cdot TE$$
$$k_2 = \epsilon_h \cdot r_0 \cdot E_0 \cdot TE \tag{27}$$
$$k_3 = 1 - \epsilon_h$$

The three terms inside the brackets, weighted by coefficients $k_1$, $k_2$ and $k_3$, relate to the extravascular contribution to the BOLD signal, the intravascular contribution and the ratio of extravascular and intravascular signals respectively.

## 14.4 Summary and implementation

The implementation of haemodynamic model in SPM is split into two parts: the rCBF / Balloon model (**spm_fx_fmri.m**) and the BOLD signal model (**spm_gx_fmri.m**). They include five hidden states: the net neural activity $z$, vasoactive signal $S$, log rCBF $\ln(f_{in})$, log venous volume $\ln(v)$ and log dHb $\ln(q)$. The logs of these states are taken in order to enforce positivity constraints, requiring the differential equations to be supplemented accordingly:

$$\frac{d \ln(f_{in})}{dt} = \frac{\dot{f}_{in}}{f_{in}}$$

$$\frac{d \ln(v)}{dt} = \frac{\dot{v}}{v} \tag{28}$$

$$\frac{d \ln(q)}{dt} = \frac{\dot{q}}{q}$$

There are 10 parameters overall, labelled in Figure A.2. To keep estimation of the model tractable, only three of these are estimated from the fMRI data, and the log of their values are estimated to ensure positivity: the rate of signal decay $\ln(\kappa)$ from the rCBF model, the transit time $\ln(\tau_h)$ from the balloon model, and the ratio of intra- to extra-vascular signals $\ln(\epsilon_h)$ from the BOLD signal model. The priors for these parameters are listed in Table 2.

# 16 Tables

## 16.1 Table 1: Symbols

| Variable | Dimension | Units | Meaning |
|---|---|---|---|
| $A$ | $R \times R$ | Hz | Effective connectivity (average or baseline) |
| $A_E$ | $R \times R$ | Hzs | Extrinsic average or baseline effective connectivity |
| $A_I$ | $R \times R$ | - | Log scaling parameters on average or baseline intrinsic connections |
| $\alpha$ | $1 \times 1$ | - | Grubb's exponent (stiffness of blood vessels) |
| $B^{(k)}$ | $R \times R$ | Hz | Modulatory input parameters for condition $k$ |
| $B_E^{(k)}$ | $R \times R$ | Hz | Modulation of extrinsic connections by condition $k$ |
| $B_I^{(k)}$ | $R \times R$ | - | Log scaling parameters on modulation of intrinsic connections by condition $k$ |
| $B_0$ | $C_0 \times R$ | - | Parameters for null effects |
| $C$ | $R \times J$ | Hz | Driving input parameters |
| $C_0$ | $1 \times 1$ | - | Number of first level covariates of no interest |
| $E_0$ | $1 \times 1$ | - | Resting oxygen extraction fraction |
| $\epsilon$ | $V \times 1$ | - | Observation noise |
| $\epsilon_h$ | $1 \times 1$ | - | Fraction of intravascular to extravascular signal |
| $F$ | $1 \times 1$ | Nats | Negative variational free energy for a given model |
| $f$ | - | - | Neural model |
| $f_{in}$ | $1 \times 1$ | Hz | Rate of blood inflow |
| $f_{out}$ | $1 \times 1$ | Hz | Rate of blood outflow |
| $g$ | - | - | Observation model |
| $\gamma$ | $1 \times 1$ | Hz | Rate of decay of feedback to vasodilatory signal |
| $J$ | $R \times R$ | Hz | Effective connectivity or Jacobian matrix |
| $K$ | $1 \times 1$ | - | Number of experimental conditions |
| $k_x$ | $1 \times 1$ | - | Coefficient within the BOLD signal model |
| $\kappa$ | $1 \times 1$ | Hz | Rate of vasodilatory signal decay |
| $\lambda_i$ | $1 \times 1$ | - | Log scaling parameter for covariance component $i$ |
| $P_H^{(1)}$ | $1 \times 1$ | - | Total haemodynamic parameters per DCM |
| $P_N^{(1)}$ | $1 \times 1$ | - | Total neural parameters per DCM |
| $P_\epsilon^{(1)}$ | $1 \times 1$ | - | Total observation parameters per DCM |
| $\Pi_y$ | $(T \cdot R) \times (T \cdot R)$ | - | Precision of observations (measurements) |
| $Q_i$ | $(T \cdot R) \times (T \cdot R)$ | - | Covariance component $i$ |
| $q$ | $1 \times 1$ | - | Level of deoxyhaemoglobin normalized to rest |
| $R$ | $1 \times 1$ | - | Number of modelled brain regions |
| $R^*$ | $1 \times 1$ | - | Total voxels (and timeseries) in the MRI volume |
| $R_{2E}^*$ | $1 \times 1$ | Hz | Extravascular transverse relaxation rate |
| $R_{2I}^*$ | $1 \times 1$ | Hz | Intravascular transverse relaxation rate |
| $r_0$ | $1 \times 1$ | - | Constant relating $R_{2I}^*$ to oxygen extraction rate |
| $s$ | $1 \times 1$ | - | Vasodilatory signal |
| $S$ | $1 \times 1$ | - | Modelled BOLD signal |
| $S_0$ | $1 \times 1$ | - | Modelled BOLD signal at rest |
| $S_E$ | $1 \times 1$ | - | Extravascular contribution to $S$ |
| $S_I$ | $1 \times 1$ | - | Intravascular contribution to $S$ |
| $S_{E0}$ | $1 \times 1$ | - | Extravascular effective spin density |
| $S_{I0}$ | $1 \times 1$ | - | Intravascular effective spin density |
| $\Sigma_y$ | $(T \cdot R) \times (T \cdot R)$ | - | Covariance of the observations (measurements) |

| $T$ | $1 \times 1$ | - | Total time points in the inputs $U$ |
|---|---|---|---|
| $TE$ | $1 \times 1$ | Secs | Echo time |
| $\tau_h$ | $1 \times 1$ | Secs | Haemodynamic transit time |
| $\vartheta_0$ | | - | Frequency offset - outer surface of magnetized values |
| $\theta^{(h)}$ | $P_H^{(1)} \times 1$ | - | All first level haemodynamic parameters |
| $\theta^{(n)}$ | $P_N^{(1)} \times 1$ | - | All first level neural parameters |
| $\theta^{(\epsilon)}$ | $P_\epsilon^{(1)} \times 1$ | - | All first level observation parameters |
| $U$ | $T \times K$ | - | All experimental inputs |
| $u(t)$ | $J \times 1$ | - | All experimental inputs at time $t$ |
| $u_k(t)$ | $1 \times 1$ | - | Experimental input by condition $k$ at time $t$ |
| $\vartheta_0$ | $1 \times 1$ | Hz | Frequency offset at the outer surface of magnetised vessels |
| $V$ | $1 \times 1$ | - | Total measurements (volumes) per subject |
| $v$ | $1 \times 1$ | - | Blood volume normalized to rest |
| $V_0$ | | - | Resting blood volume fraction |
| $V_1$ | | - | Blood volume fraction following neural activity |
| $V_h$ | $1 \times 1$ | - | Fraction of intravascular blood volume |
| $X_0$ | $V \times C_0$ | - | Design matrix for null effects |
| $Y$ | $V \times R$ | - | Observed timeseries from all regions of interest |
| $\hat{Y}$ | $V \times R^*$ | - | All timeseries from the acquired MRI volume |
| $z$ | $R \times 1$ | - | Neural activity in each region |

## 16.2 Table 2: Free parameters and their priors

| Name | Parametrization † | Prior expectation | Prior variance (uncertainty) | 90% CI |
|---|---|---|---|---|
| $A_E$ | $A_E$ | 0 | 1/64 | [-0.21 0.21] |
| $A_I$ | $-0.5\text{Hz} \cdot \exp(A_I)$ | 0 | 1/64 | [-0.21 0.21] |
| $B_E$ | $B_E$ | 0 | 1 | [-1.65 1.65] |
| $B_I^{(k)}$ | $-0.5\text{Hz} \cdot \exp\left(B_I^{(k)}\right)$ | 0 | 1 | [-1.65 1.65] |
| $C$ | $\frac{1}{16} \cdot C$ | 0 | 1 | [-1.65 1.65] |
| $\epsilon$ | $\exp(\epsilon)$ | 0 | 1/256 | [-0.10 0.10] |
| $\kappa$ | $0.64\text{Hz} \cdot \exp(\kappa)$ | 0 | 1/256 | [-0.10 0.10] |
| $\lambda_i$ | $\exp(\lambda_i)$ | 6 | 1/128 | [5.85 6.14] |
| $\tau_h$ | $2.00\text{s} \cdot \exp(\kappa)$ | 0 | 1/256 | [-0.10 0.10] |

† Log scaling parameters have no units - they are exponentiated and then multiplied by fixed default values, listed in the Parametrization column.

## 16.3 Table 3: Example subject's neural parameters

| | Parameter* | Description | Units | Expectation | Precision | Probability† |
|---|---|---|---|---|---|---|
| 1 | $A_{I\,11}$ | Self-connection on lvF | None | -0.16 | 66.94 | 0.91 |
| 2 | $A_{I\,22}$ | Self-connection on ldF | None | -0.04 | 68.64 | 0.62 |
| 3 | $A_{I\,33}$ | Self-connection on rvF | None | -0.04 | 75.39 | 0.62 |
| 4 | $A_{I\,44}$ | Self-connection on rdF | None | -0.18 | 93.87 | 0.96 |
| 5 | $A_{E\,21}$ | lvF → ldF | Hz | 0.42 | 233.16 | 1.00 |
| 6 | $A_{E\,31}$ | lvF → rvF | Hz | 0.06 | 406.70 | 0.88 |
| 7 | $A_{E\,12}$ | ldF → lvF | Hz | -0.02 | 291.40 | 0.58 |
| 8 | $A_{E\,42}$ | ldF → rdF | Hz | 0.57 | 145.30 | 1.00 |
| 9 | $A_{E\,13}$ | rvF → lvF | Hz | 0.43 | 149.48 | 1.00 |
| 10 | $A_{E\,43}$ | rvF → rdF | Hz | 0.10 | 102.21 | 0.86 |
| 11 | $A_{E\,24}$ | rdF → ldF | Hz | -0.03 | 483.41 | 0.73 |
| 12 | $A_{E\,34}$ | rdF → rvF | Hz | -0.21 | 858.90 | 1.00 |
| 13 | $B_{I\,11}^{(2)}$ | Pictures on lvF self | None | -0.47 | 41.73 | 1.00 |
| 14 | $B_{I\,22}^{(2)}$ | Pictures on ldF self | None | 2.12 | 3.52 | 1.00 |
| 15 | $B_{I\,33}^{(2)}$ | Pictures on rvF self | None | 0.13 | 16.78 | 0.70 |
| 16 | $B_{I\,44}^{(2)}$ | Pictures on rdF self | None | -0.16 | 19.21 | 0.68 |
| 17 | $B_{I\,11}^{(3)}$ | Words on lvF self | None | 2.80 | 1.98 | 1.00 |
| 18 | $B_{I\,22}^{(3)}$ | Words on ldF self | None | 0.27 | 9.98 | 0.81 |
| 19 | $B_{I\,33}^{(3)}$ | Words on rvF self | None | 0.24 | 6.40 | 0.73 |
| 20 | $B_{I\,44}^{(3)}$ | Words on rdF self | None | 0.11 | 13.41 | 0.71 |
| 21 | $C_{11}$ | Driving: task on lvF | Hz | -0.07 | 910.27 | 0.99 |
| 22 | $C_{21}$ | Driving: task on ldF | Hz | 0.10 | 909.84 | 1.00 |
| 23 | $C_{31}$ | Driving: task on rvF | Hz | 0.26 | 811.03 | 1.00 |
| 24 | $C_{41}$ | Driving: task on rdF | Hz | 0.08 | 474.01 | 0.96 |

*Region names: 1=lvF, 2=ldF, 3=rvF, 4=rdF. Condition names (superscript on matrix $B_I$): 2=Pictures, 3=Words. †Probability that the posterior estimate of the parameter is not zero. For a parameter with marginal posterior density $N(\mu, \sigma^2)$ this is given by $1 - NCDF(abs(\mu), \sigma^2)$, where NCDF is the normal cumulative density function.